\documentclass[12pt,preprint]{aastex}

\begin{document}

\title{HST and Optical Data Reveal White Dwarf Cooling, Spin and Periodicities in GW Librae 3-4 Years
after Outburst\altaffilmark{1}}

\author{Paula Szkody\altaffilmark{2,3,4},
Anjum S. Mukadam\altaffilmark{2},
Boris T. G\"ansicke\altaffilmark{5},
Arne Henden\altaffilmark{6},
Edward M. Sion\altaffilmark{7},
Dean Townsley\altaffilmark{8},
Paul Chote\altaffilmark{9,4},
Diane Harmer\altaffilmark{10},
Eric J. Harpe\altaffilmark{11},
J. J. Hermes\altaffilmark{12},
Denis J. Sullivan\altaffilmark{9,4},
D. E. Winget\altaffilmark{12}}
\altaffiltext{1}{Based on observations
made with the NASA/ESA Hubble Space
Telescope, obtained at the Space Telescope Science Institute, which is
operated by the Association of Universities for Research in Astronomy, Inc.,
(AURA)
under NASA contract NAS 5-26555, with the Apache Point Observatory 3.5m
telescope which is owned and operated by the Astrophysical Research
Consortium}
\altaffiltext{2}{Department of Astronomy, University of Washington, Seattle, WA 98195; szkody@astro.washington.edu, mukadam@astro.washington.edu}
\altaffiltext{3}{Visiting Astronomer, Kitt Peak National Observatory, National Optical Astronomy Observatory, which is operated by the Association of Universities for Research in Astronomy (AURA) under cooperative agreement with the National Science Foundation}
\altaffiltext{4}{Visiting Astronomer, Mt. John University Observatory, operated
by the Department of Physics \& Astronomy, University of Canterbury, NZ}
\altaffiltext{5}{Department of Physics, University of Warwick, Coventry
CV4 7AL, UK; boris.gaensicke@warwick.ac.uk }
\altaffiltext{6}{AAVSO, 49 Bay State Road, Cambridge, MA 02138; arne@aavso.org}
\altaffiltext{7}{Department of Astronomy \& Astrophysics, Villanova University,
Villanova, PA 19085; edward.sion@villanova.edu}
\altaffiltext{8}{Department of Physis \& Astronomy, University of Alabama,
Tuscaloosa, AL 35487; Dean.M.Townsley@ua.edu}
\altaffiltext{9}{School of Chemical \& Physical Sciences, Victoria University of Wellington, New Zealand; denis.sullivan@vuw.ac.nz, paul.chote@vuw.ac.nz}
\altaffiltext{10}{National Optical Astronomy Observatories, 950 North Cherry Avenue, Tucson, AZ 85726; diharmer@noao.edu}
\altaffiltext{11}{Heritage High School, 7825 NE 130th Avenue, Vancouver, WA 98682; Eric.Harpe@evergreenps.org}
\altaffiltext{12}{Department of Astronomy, University of Texas, Austin, TX 78712; dew@astro.as.utexas.edu}

\begin{abstract}
Since the large amplitude 2007 outburst which heated its accreting, pulsating
white dwarf, the dwarf
nova system GW Librae has been cooling to its quiescent
temperature. Our Hubble Space Telescope ultraviolet spectra combined
with ground-based optical coverage during the 3rd and 4th year after
outburst show that the fluxes and temperatures are still
higher than quiescence (T=19,700K and 17,300K vs 16,000K pre-outburst
for a log g=8.7 and d=100 pc). The K$_{\mathrm{wd}}$ of 7.6$\pm$0.8 km s$^{-1}$ 
determined from the CI$\lambda$1463
absorption line, as well as the gravitational
redshift implies a white dwarf mass of
0.79$\pm$0.08 M$_{\odot}$.  The widths of the UV lines imply a white
dwarf rotation velocity vsin i of 40 km s$^{-1}$ and a spin period of 209 s
(for an inclination of 11 deg and a white dwarf
radius of 7$\times$10$^{8}$cm).
Light curves produced from the UV spectra in both years 
show a prominent multiplet near 290 s, with higher amplitude 
in the UV compared to
the optical, and increased amplitude in 2011 vs 2010.
As the presence of this set of periods is intermittent in the optical
on weekly timescales, it
is unclear how this relates to the non-radial pulsations evident during
quiescence.
\end{abstract}


\section{Introduction}

The dwarf nova GW Librae has undergone two very large amplitude outbursts,
the first during its discovery in 1983 (Gonzalez \& Maza 1983) and the second in
April 2007 (Templeton et al. 2007). While it was V=17.0 magnitude (Thorstensen et al. 2002)
during its long quiescence, it reached 8th magnitude at outburst. It's very
short orbital period of 76.78 min (Thorstensen et al. 2002) and long outburst
recurrence time are consistent with very low accretion rate dwarf novae (Howell,
Szkody \& Cannizzo 1995). This low accretion at quiescence 
allows a view of the white
dwarf, which was found to show non-radial pulsations at 648, 376 and 236 s 
(Warner \& van Zyl 1998, van Zyl et al. 2000, 2004).
Ultraviolet observations  with the Space Telescope Imaging Spectrograph (STIS)
showed the same pulse periods with amplitudes 6-17 times higher than the
optical (Szkody et al. 2002), consistent with limb-darkening effects
in the atmospheres of stellar pulsators (Robinson et al. 1995). The 2002 UV
spectrum revealed a hot white dwarf at $\sim$15,000K for log g=8.0; this
temperature places GW Lib near the blue edge of
the instability strip for accreting pulsating white dwarfs (Szkody et al. 2010)
prior to its outburst. Townsley et al. (2004) used the pulsation
periods and the UV data to estimate a high mass of 1.02M$_{\odot}$ 
for the white dwarf.

The 2007 outburst was well-studied and provided several interesting
avenues for determining various parameters of GW Lib. A superhump
present soon after outburst (Kato et al. 2008), used with
the orbital period and the empirical relation of Patterson et al. (2005),
gave a mass ratio M$_{2}$/M$_{1}$=0.06. A narrow emission component
from the irradiated donor star provided a K velocity (82$\pm$5 km s$^{-1}$) 
and systemic velocity (-15$\pm$5 km s$^{-1}$) for the system (van Spaandonk et
al. 2010a). As the white dwarfs in dwarf novae are known to be heated
by their outbursts (Long et al. 1994, Sion et al. 1998, Piro et al. 2005, Godon et al. 2006),
GW Lib presents the unique opportunity to determine the effect of the
outburst on the interior of the white dwarf by following the pulsations.
The expectation is that the pulsations will stop as the heating causes
the white dwarf to move out of the instability strip, and then resume,
possibly with shorter periods than previously observed during quiescence,
as the white dwarf cools and re-enters the blue edge of the instability strip.
A shorter period post-outburst can arise for two reasons: a
higher surface temperature may excite shorter-period modes (Arras et al 2007) or
the higher temperature surface layer can increase the local buoyancy
(Townsley et al 2004).  However, a computation of the effect of a heated outer
layer on the g-mode spectrum of a white dwarf, let alone a rapidly rotating
one, has not yet been performed.  It is possible that the low-order
eigenmodes are affected in a non-obvious way. 
If the white dwarfs are like the H-rich, DA white dwarf pulsators (ZZ Ceti), where
the thermal timescale at
the base of the convection zone determines which periods will be excited
(Montgomery 2005), the periods observed should move from short to longer
values as the star cools and the convection zone moves deeper into the star.
This scenario is consistent with the observed 
results for ZZ Ceti
 pulsators, where those
closer to the red edge of the instability strip have longer periods than
the hotter pulsators near the blue edge (Clemens 1993, Mukadam et al. 2006).
The re-appearance and subsequent evolution of the pulsation spectrum allows
an important contstraint on the depth of heating that occurs during the outburst. As
the mass transfer in accreting, pulsating white dwarfs likely results in
different compositions,
increased rotation and increased heating compared to ZZ Ceti stars,
their study will help us grasp how these parameters affect the non-radial pulsations.

Follow-up
near-UV and optical photometry for the 3 yrs following the outburst of
GW Lib (Copperwheat
et al. 2009, Schwieterman et al. 2010, Bullock et al. 2011, Vican et al. 2011)
showed decreasing temperatures with a long period near 4 hrs and a quasi-period at 19 min (evident for 2-4 months) attributed to disk phenomena but no
evidence for the return of the non-radial pulsations. The optical magnitude
remained at about 0.5 mag above its quiescent level.

We accomplished {\it Hubble Space Telescope (HST)} ultraviolet 
observations in 2010-2011 to follow the final
return of GW Lib to its quiescent temperature. This paper presents
our results together with ground-based observations conducted during
this same interval.

\section{Observations}

The observations from space and ground took place during 2010-2011, the third
and fourth years after the large amplitude dwarf nova outburst of GW Lib.
 
\subsection{HST ultraviolet spectra}

Two sets of ultraviolet spectra were obtained with the Cosmic Origins Spectrograph (COS).
Observations during five HST orbits took place on 2010 March 9; the first four
 with the G160M grating and the last one with G140L. The G160M observations covered 
a wavelength range of 1405-1775\AA\ with a resolution of $\sim$0.07\AA, while the
G140L has a wider bandpass (1130-2000\AA) with 
lower resolution ($\sim$0.75\AA). The wavelength ranges that provided useful
data for GW Lib were 1388-1558 and 1579-1748\AA\ for G160M and 1130-1860\AA\ for
G140L.

The second set of spectra were obtained
 during two HST orbits on 2011 April 9, using
the G140L grating for both orbits. The time-tag data were analyzed with PyRAF
routines from the STSDAS task package hstcos (version 3.14). The summed 
spectrum from each grating was extracted with a series of
widths to optimize the S/N. For the G160M grating with setting 1577, 
an extraction width of 27 pixels was used as opposed to the default
value of 35 for the primary science aperture (PSA). The optimum extraction
for the G140L grating setting
of 1105 was 41 pixels versus the default value of 57.
 The 4 orbits of G160M
data were phased on the orbital period of GW Lib and binned into 10 phase bins. 
These data were used to study the orbital velocity of the white dwarf and
its rotation.

Light
curves were created from all the spectra by summing the fluxes over all useful
wavelengths on several short timescales (3-30s)
to search for variability on orbital and pulsation timescales. 
Light curves were created after deleting the emission lines from the data in
order to reduce the contribution from the accretion disk, and focus on the
white dwarf variability.
The light curves were
divided by the mean and then one was subtracted to place them on a
fractional amplitude scale which was used for DFT analysis. 
 A log of the HST observations is given
in Table 1.
 
\subsection{Ground-based optical photometry}

Optical photometry was planned close to the times of HST observation, both
to ensure that GW Lib was close to its quiescent brightness (required by HST)
and to monitor the return of pulsations. AAVSO observations provided
brightness measurements before, during and following the HST times and
are available through their archive{\footnote{http://www.aavso.org/data-access}}.
These data showed that GW Lib was at a V magnitude near 16.5 in 2010
and 16.75 in 2011.
Four nights of time-series photometry on larger telescopes with broad-band
blue filters were
obtained in the 2010 season and 16 nights during 2011. Six different
telescopes were used during these times with broadband blue bandpass filters. 
The 2.1m and 0.9m telescopes at
McDonald Obervatory (MO) were used with BG40 filters; the 2.1m with the
Argos time-series CCD (Nather \& Mukadam 2004), and the 0.9m with a
comparable CCD called Raptor. The 3.5m at Apache Point Observatory (APO) also
employed the same filter and a similar frame-transfer CCD which is called Agile 
(Mukadam et al. 2011a) and the
Mt.\ John University Observatory (MJUO) 1m telescope in New Zealand used the same CCD camera and
filter as
in Agile and is called Puoko-Nui.
The Las Cumbres Observatory
Global Telescope network 2m Faulkes Telescope South (FTS) provided
observations with a
Fairchild CCD and Bessell B filter. The Kitt Peak National Observatory
(KPNO) 2.1m telescope with the STA2 CCD and BG39 filter was used for
short runs on several nights. A summary of these observations is also given in
Table 1. The optical data were converted to fractional amplitude in the
same way as the HST data and used to compute DFTs.
 
\subsection{Optical Spectra}
One optical spectrum was obtained in 2010 May 5 with the Dual Imaging
Spectrograph (DIS), using the high resolution grating (2\AA) and a 1.5
arcsec slit. One 600 sec exposure produced a blue spectrum from
4000-5150\AA\ and a red spectrum from 6275-7375\AA.

\section{Results}

Due to the larger wavelength coverage, the lower resolution G140L spectra
were used for temperature determination through fits to white dwarf
models and a subsequent cooling curve. The higher resolution G160M spectra were used for line measurements
to construct a radial velocity curve for the white dwarf and line width
fitting to determine the rotation of the white dwarf. All spectra were
used to search for pulsations through the creation of light curves. These
results are described in detail below.

\subsection{Cooling Curve}

We have three available UV spectra of GW Lib with which to determine the white
dwarf temperature: the four orbits at quiescence 
with STIS (resolution
of 1.2\AA; Szkody et al. 2002) from 2002 January 17; the one orbit of
COS G140L at 3 years past outburst on 2010 March 11 and the 2 orbits
with COS G140L on 2011 April 9 at 4 years past outburst. Figure 1
shows these 3 spectra on the same scale, with the COS data binned to
the slightly lower resolution of the STIS data. The increased temperature
and flux from the post outburst data are immediately apparent, as is the
fact that even at 4 years past outburst, the white dwarf remains $\sim$1000K 
hotter than its quiescent level. 

Since distance and gravity and metal abundance 
affect the fitting result, we treated all
3 spectra by using the distance of 100 pc determined from parallax by 
Thorstensen (2003), adopting a log g=8.7, corresponding to a mass
near 1M$_{\odot}$ and using a metal abundance of 0.1 times the solar values. 
The procedure to determine the white dwarf temperature
was similar to that described in G\"ansicke et al. (2005). The contribution
of the disk was estimated as a black-body component that accounts for the residual
flux in the core of Ly$\alpha$. In all cases, this component contributes
only a few percent, and past experience has shown it does not matter to
the temperature fit if a black-body or power law is used over the short
wavelength range. The geocoronal Ly$\alpha$ emission is removed from
the fit and the CIV, SIV and HeII emission lines are treated with broad Gaussians. Then
the spectrum is matched with a grid of white dwarf models (Hubeny \& Lanz
1995) to find the best fit. The resulting fits are listed in Table 2
and shown in Figure 2. For our fixed log g, the uncertainties in the
temperature fits are about 200K. Decreasing log g by 0.5 dex decreases
the best-fit temperatures by $\sim$1000K and increases the distance
as these parameters are tied together. The metal abundance will also
have a small effect.

To determine a temperature estimate closer to the outburst time, we
used the fit (25,000K) to an optical spectrum obtained on 2008 April 17 
(one year
after outburst) shown in Bullock et al. (2011). However, since the optical is much more
contaminated by the disk, this value is not as well-determined as the UV
temperatures. Figure 3 shows the flux decrease in the 
blue and red optical spectra obtained at APO on 2010 May 5 in comparison to the 2008 
blue optical spectrum. Since the APO spectra were obtained at high airmass compared
to the 2008 spectrum (from CTIO), some of the blue decrease may be the result of
extinction or refraction. Thus, we only used the 2008 optical and the UV data from
2010 and 2011 for a cooling curve.
 The resulting temperatures from the available data are shown in Figure
4 along with a a 1D quasi-static evolutionary simulation (Sion 
1995, Godon et al. 2006) of the heating and subsequent cooling of GW Lib's 
white dwarf in response to the 2007 outburst. In this simulation, accretion 
at a high rate was switched on for 23 days to simulate the mass deposition 
and heating of the outburst. The simulation was carried out for a 1M$_{\odot}$ 
white dwarf. The best agreement with the empirical temperature decline was 
obtained for an outburst accretion rate of 
5$\times$10$^{-8}$ M$_{\odot}$ yr$^{-1}$. The curve for 
GW Lib is very similar in shape to that found for WZ Sge (Godon et al. 2006)
but the cooling rate of GW Lib's white dwarf appears to be slower than the 
WZ Sge 
degenerate despite having a comparable mass (0.85M$_{\odot}$; Steeghs et al. 2007). WZ Sge accreted at a
high (outburst) rate for 52  days while GW Lib accreted at its outburst rate for
23 days. However, WZ Sge has a cooler quiescent white dwarf (13,500K vs 16000K)
and its outburst amplitude was only 7 mag while GW Lib was 9 mag.
Differences in outburst accreted mass, viewing angle, as well as long term
accretion rates may account for
the different temperatures in the two systems. 

\subsection{Radial Velocity Curves}

The time-tag spectra from the four orbits of optimized high resolution G160M data 
obtained in 2010 were binned into 10 phases
using the orbital period and taking the start of the observations as
the arbitrary zero phase. Absorption lines of CI$\lambda$1463, 
SiII$\lambda$ 1526,1533
and Al$\lambda$1670 were deemed strong enough for useful velocity determination
throughout the orbit. These lines were measured singly with 
IRAF\footnote{IRAF (Image
 Reduction and Analysis
Facility) is distributed by the National Optical Astronomy Observatories, which
are operated by AURA,
Inc., under cooperative agreement with the National Science Foundation.}
routines ``e'' which determines a centroid and ``k'' which uses a gaussian
fit. Since the SiII 1526 line is blended with an ISM component, a procedure that fit the photospheric lines and
interstellar component with 3 Gaussians and only allowed the photospheric
lines to move together was tried. This latter procedure showed a mean
statistical error on the velocities of 3.6 km s$^{-1}$ but no apparent
overall radial velocity variation. However, Gaussian fits from IRAF on
the CI line alone produced a noticeable velocity curve. Figure 5 shows the
measurements from the single fits to the three lines (SiII is only the 1533\AA\ component) along
with the best fit sine-curve to the velocities to determine 
$\gamma$ (systemic velocity) and K (semi-amplitude) using the fixed
known period of GW Lib. The solutions are listed in Table 3 along with
the total $\sigma$ of the fit. While only the CI solution (K=7.6$\pm$0.08 km s$^{-1}$) appears
viable, the fits to the other 2 lines have similar shapes as a function of phase. It is
 apparent that K$_{\mathrm{wd}}$ is very small, much smaller
than the past estimates from optical emission lines (38$\pm$3 km s$^{-1}$
from Thorstensen et al. 2002 and 19.2$\pm$5.3 km s$^{-1}$ from van Spaandonk et al. 2010a). Using a narrow CaII line on the secondary that appeared near outburst
to determine K$_{\mathrm{sec}}$ in combination with a superhump period excess to
determine the mass ratio q,  van Spaandonk et al. (2010b) determined
K$_{\mathrm{wd}}$ of 6.25$\pm$0.4 km s$^{-1}$. This value is consistent with the direct
measurement from the CI line. It is also consistent with the center
of symmetry of the disk (6$\pm$5 km s$^{-1}$) determined from the Doppler map 
of the broad CaII emission (van Spaandonk et al. 2010b). Using K$_{\mathrm{wd}}$ from
the CI line and K$_{\mathrm{sec}}$ from van Spaandonk et al. (2010b) implies a
white dwarf mass of 0.79$\pm$0.08 M$_{\odot}$. This value is consistent with
the average mass for the white dwarfs in cataclysmic variables, and higher than the masses of single white dwarfs and those in
pre-cataclysmic binaries (Zorotovic et al. 2011).

The sine-fit to the velocity curve of the CI line can also be used to
determine the gravitational redshift of the white dwarf and another
estimate for its mass. Van Spaandonk et al (2010b) give an extensive
discussion of corrections for systemic velocity as derived from the
donor star and CaII from the disk; we use their value of -18.1$\pm$2.0
together with our velocity of 29.3$\pm$0.1 for CI (Table 3) to determine
v$_{\mathrm{grav}}$ (WD) = 47$\pm$2 km s$^{-1}$. This value is similar to that
found by van Spaandonk et al (2010b) from the weak MgII triplet and
consistent with the mass from the K velocity above. 

The broad CIV$\lambda$1550 emission line was also measured and shows a prominent
variation that is offset from the absorption lines. Figure 6 shows
the measurements and the sine-fit that are listed in Table 3. While
the phase shifts of the C, Si and Al lines are all the same (even though
the solutions are poor for Si and Al), the CIV emission is offset by 0.3
phase. Van Spaandonk et al. (2010a) pointed out a phase offset of 0.7 phase
between the CaII narrow emission from the secondary and the CaII broad emission
from the accretion disk. Since we don't know the absolute phasing,
it is not clear if these offsets are related but it appears that the CIV 
emission
cannot be located in a symmetrical region close to the white dwarf.

\subsection{White Dwarf Rotation}

The high resolution G160M data also allow an estimate of the rotation
of the white dwarf, using the widths of the absorption lines. The white dwarf
models were broadened with several rotation rates and compared to the
spectrum (Figure 7). The plot shows various rotation rates (20, 50 and
87 km s$^{-1}$ in red, green and magenta respectively) for the region of
the SiII lines, using a white dwarf 
model with T=17,500K, log g=9.0 and 0.1 solar abundances. While the v$\sin i$ 
changes slightly with metal abundance
and temperature (G\"ansicke et al. 2005), it is clear that the best fit is somewhere near 40 km s$^{-1}$.
For an inclination of 11 deg (van Spaandonk et al.2010b), the rotation 
velocity = 210 km s$^{-1}$ and 
with a white dwarf radius of 7$\times$10$^{8}$ cm, the rotation period
is 209 s. Van Spaandonk et al. (2010b) found v$\sin i$ to be 
87$\pm$3 km s$^{-1}$ using a weak triplet MgII absorption line in the
optical (yielding a rotation period of 97 s for the same parameters of
inclination and radius). The deeper, unblended lines apparent in the UV 
combined with the higher S/N provide a more reliable value for this 
parameter. The rotation of GW Lib is within the range determined for a few
other dwarf novae (for a similar radius, the observed v$\sin i$ implies
a period of 63 s for VW Hyi, 110-114 s for WZ Sge and 400-800 s for U Gem;
Sion et al. 1995, Long et al. 2003, Sion  et al. 1994).  
Our result confirms the fast rotation of GW lib compared to
single white dwarfs and rules out the spin as the cause of the fine structure 
with $\mu$Hz spacing apparent in its
pulsation spectrum at quiescence (van Zyl et al. 2004). This
 fine structure is also
apparent in another accreting pulsator SDSS1610-0102 (Mukadam et al. 2010), 
which has no adequate resolution UV spectrum to confirm its rotation. Note that the spin period of GW Lib does not show
up in the quiescent power spectrum, as it does in the pulsator V455 And
(Araujo-Betancor et al. 2005), so the spectral line fits provide the only
means to determine the rotation.

\subsection{Light Curves}

The UV and optical light curves throughout 2010 and 2011 were treated
in a similar manner as in Mukadam et al. (2010). The DFT's were
computed for each light curve and the amount of white noise in the
light curve was determined empirically by using a shuffling technique
(see Kepler 1993). All the best-fit frequencies were initially subtracted
to obtain a pre-whitened light curve. Preserving the time column of this
light curve, the corresponding intensities were shuffled to destroy any
coherent signal while keeping the time sampling intact. A DFT of the shuffled
light curve was then computed and its average amplitude was taken as the
1$\sigma$ limit of white noise. After shuffling the light curve 10 times,
the corresponding values for white noise were averaged to determine a reliable
3$\sigma$ limit.
Table 4 lists the 3$\sigma$ limits for all nights and the periods above this 
value which were present on at least 3 nights.
 The UV data have the advantage of larger amplitudes
of pulsation due to limb-darkening effects (Robinson et al. 1995; Szkody
et al. 2002). However, the HST data are constrained to at most 5 consecutive
orbits with large time gaps between orbits. The optical runs could be longer
and gap-free but the southern declination of GW Lib enabled long runs
only from the southern hemisphere observatory Mt. John. The 2010 March
14 data had only 36 useful images so no DFT was computed. A long period
near 2500 s (41.7 min) appears in a good number of datasets, as well as
a period near 5100 s (85 min). These periods
do not appear to be related to the orbital period or even the superhump
period seen after outburst. They could be caused by some structure in the
disk which has not yet returned to its quiescent state.

The first significant evidence of a short period that could be
related to the return of pulsations comes from the COS data of
2010. All five orbits (4 of G160M and 1 of G140L) extracted with 3 s
exposures are shown in Figure 8. The fractional intensity variation of
each orbit is shown at the top, with a view of the entire DFT in the
middle and an expanded view of the section around the prominent peak
at 0.0035 Hz in the bottom plot. There are 4 closely-spaced periods at
266, 282, 289, and 292 s, with the 292 s period having the highest
amplitude (Table 4). The optical data from the
McDonald 0.9m telescope that is simultaneous with
part of this time (Table 1) show a period (284 s) within this frequency
range
(Figure 9). The shorter run on the
2.1 m telescope the next day (with a stricter 3$\sigma$) limit does not
show any significant feature. The 290 s period appears to be formed of
multiple components
but the short stretches of data may not be resolving it properly. The
266 s period may be a separate mode.

The two COS orbits in 2011 reveal a periodicity
in this same vicinity at 293 s (Figure 10), with an amplitude more than twice as
large as the 2010 data. The shorter 
time span (2 orbits in 2011 versus the 4 in 2010) does not
permit the resolution to determine if this feature is also split as in 2010, 
but the peak is unmistakable.
Figure 11 shows a comparison of the G140L data from both years on the
top and a comparison of the DFTs on the bottom. The closest optical data sets
to the HST also show significant power near this frequency (Table 4).
Figure 12 shows the DFTs for 2011 March-April from the MJUO and APO telescopes.

Further optical data (shorter runs from KPNO in 2011 May-June) and longer
runs from Mt. John in 2011 July-Aug show an intermittent
 large amplitude periodicity near 290 s is present on some nights
 (Figures 13 and 14 and Table 4), but absent on others. In
addition, as both the HST and optical data show, the period is not
at exactly the same frequency when it is evident. Other intermittent periods
between 300-320 s are also apparent. The presence of
different periods at different times in ZZ Ceti stars is a known phenomena
of amplitude modulation which is not well understood (Kleinman et al. 1998).
The appearance of different pulsation frequencies was visible in GW Lib
at quiescence (van Zyl et al. 2004) and has been seen in the newly
discovered accreting pulsators SDSSJ1457+51 and BW Scl at quiescence
(Uthas et al. 2012). On the other hand, the pulsations in SDSS0745+45
were observed to occur at the identical frequencies 3 yrs after outburst
as at quiecence (Mukadam et al. 2011). Since the pulsation spectrum
of GW Lib 3-4 yrs after outburst (closely-spaced periods near 290 s) compared
to quiescence (3 periods at 237, 376 and 646 s), is very different, 
it is difficult to
determine whether or not this is the actual return of pulsation. Figure 15
shows a comparison of the light curves and periods from the STIS quiescent
data to the COS post-outburst data. Generally,
the hotter ZZ Cet stars show the shortest pulsation periods, so shorter
periods than quiescence are expected.

\section{Conclusions}

Ultraviolet and optical monitoring of GW Lib throughout the 3-4 years following
its large amplitude outburst has revealed the following information:

\begin{itemize}
\item The temperature of the white dwarf continued to decline from 19,700K
at 3 years to 17,300K at 4 years past outburst, but remained above its
quiescent temperature of 16,000K (for a 1 M$_{\odot}$ white dwarf). During
this time, the optical magnitude remained 0.3-0.5 mag above quiescence as well.

\item The best fit evolution simulation for the available temperatures
implies an outburst accretion rate of 5$\times$10$^{-8}$ M$_{\odot}$ yr$^{-1}$ with
the white dwarf reaching a peak temperature of $\sim$50,000K.

\item The motion of the UV absorption lines throughout the orbit is
very small, with the best measurements yielding a K$_{\mathrm{wd}}$ of 7.6$\pm$0.8 km s$^{-1}$, consistent with a relatively high mass white dwarf.

\item From the UV lines, the gravitational redshift of the white dwarf is 
determined to be 47 km s$^{-1}$, similar to the value determined by
van Spaandonk et al. (2010b), implying a white dwarf mass of 0.8 M$_{\odot}$.

\item The fit to the pronounced, resolved UV lines of SiII indicates a
rotation velocity of 40 km s$^{-1}$ for the white dwarf, implying a
spin period of 209 s for a radius of 7000 km and an inclination of 11 deg.

\item At both 3 and 4 years past outburst, the UV data show closely-spaced periods
near 290 s which appear to be a multiplet. While these periods show increased amplitudes in the UV over
the optical (factor of $\sim$5), consistent with limb-darkening expectations
for non-radial pulsations, and double in amplitude from 2010 to 2011, 
the periods are intermittent 
and very different from the pulsation spectrum seen at quiescence. It
remains to be seen if this is indeed the return of the pulsations.

\end{itemize}

Our results strengthen past ideas that a large amplitude outburst can
heat the white dwarf and result in cooling times of years (Sion 1995, 
Godon et al. 2006), that
the mean mass of the white dwarfs in cataclysmic variables is larger than
that for single white dwarfs (Zorotovic et al. 2011), and that they are
rotating much faster than their single counterparts, although not near
breakup velocity. However, the results on the non-radial pulsations are
not as clear. If the period near 290 s is the return of the pulsations,
its multiplet structure, which cannot be due to a slow spin of the white
dwarf, and its intermittent and changing frequency remain a puzzle.
As GW Lib has still not reached its quiescent temperature, further monitoring 
will still be required to determine how the interior of the white dwarf has 
reacted to
its outburst. The evolution of the pulsation spectrum is unlike that evident
in the accreting pulsator SDSS0745+45, which returned to its quiescent
pulsation spectrum within 3.3 yrs (Mukadam et al. 2011b), but then turned off 
during later
observations (Mukadam, in prep), or of SDSS0804+51, in which
pulsations appeared one year after outburst (Pavlenko 2009, Pavlenko et al. 2011), but were
not apparent later during quiescence. 
It remains unclear if the pulsations in these accreting systems can
turn on and off due to small accretion changes as well as from large
outbursts which heat the white dwarf dramatically. Long term monitoring
programs on dedicated stars will be required to sort out all the details. 

\acknowledgements
We gratefully acknowledge Elizabeth Waagen, Matthew Templeton, and 
the observers of the AAVSO for their efforts in
monitoring the brightness state of GW Lib that allowed the HST observations
to proceed, especially Bob Koff, Peter Lake, Mike Linnolt, Hazel McGee, Mike
Simonsen, Chris Stockdale and Rod Stubbings, as well as Howard Bond for
providing measurements from SMARTS. We also acknowledge the LCOGT for providing observing time.
This work was supported by NASA grants HST-GO1163.01A, HST-GO11639-01A
and HST-GO12231.01A from the Space Telescope Science Institute, which is
operated by the Association of Universities for Research in Astronomy, Inc.,
for NASA, under contract NAS 5-26555, and by NSF grant AST-1008734. 
DJS and PC acknowledge financial support from the NZ Marsden Fund. EJH acknowledges
support by the M. J. Murdock Charitable Trust.

\clearpage
\footnotesize
\begin{deluxetable}{lcclccl}
\tablewidth{0pt}
\tablecaption{Summary of Observations}
\tablehead{
\colhead{UT Date} & \colhead{Obs} & \colhead{Tel(m)} & \colhead{Instr} & 
\colhead{Filter} & \colhead{Time} & \colhead{Exp (s)}}
\startdata
2010 Mar 10 & MO & 0.9 & Raptor & BG40 & 08:25-10:28 & 30 \\
2010 Mar 11 & HST & 2.4 & COS,G160M & ... & 04:09-09:31 & time-tag \\
2010 Mar 11 & HST & 2.3 & COS,G140L & ... & 10:17-11:07 & time-tag \\
2010 Mar 11 & MO & 0.9 & Raptor & BG40 & 08:20-12:25 & 30 \\
2010 Mar 12 & MO & 2.1 & Argos & BG40 & 10:10-12:23 & 5 \\
2010 Mar 14 & LCOGT & 2.0 & Fairchild & B & 16:57-18:53 & 25 \\
2010 Mar 15 & LCOGT & 2.0 & Fairchild & B & 14:02-18:50 & 25 \\
2010 May 5 & APO & 3.5 & DIS & ... & 06:47-06:57 & 600 \\
2011 Mar 2 & MJUO & 1.0 & Puoko-Nui & BG40 & 13:10-17:23 & 20  \\
2011 Mar 4 & MJUO & 1.0 & Puoko-Nui & BG40 & 14:34-16:19 & 20  \\ 
2011 Apr 4 & APO & 3.5 & Agile & BG40 & 10:35-12:05 & 15 \\
2011 Apr 8 & APO & 3.5 & Agile & BG40 & 07:26-12:04 & 15 \\
2011 Apr 9 & HST & 2.4 & COS,G140L & ... & 14:08-16:33 & time-tag \\
2011 May 12 & KPNO & 2.1 & STA2 & BG39 & 06:40-08:44 & 30 \\
2011 May 13 & KPNO & 2.1 & STA2 & BG39 & 06:23-08:24 & 30 \\
2011 May 24 & KPNO & 2.1 & STA2 & BG39 & 05:51-08:07 & 30 \\
2011 May 25 & KPNO & 2.1 & STA2 & BG39 & 05:39-07:46 & 30 \\
2011 May 26 & KPNO & 2.1 & STA2 & BG39 & 05:50-07:41 & 30 \\
2011 Jun 9 & KPNO & 2.1 & STA2 & BG39 & 04:58-06:36 & 30 \\
2011 Jul 1 & MJUO & 1.0 & Puoko-Nui & BG40 & 07:03-12:07 & 20 \\
2011 Jul 2 & MJUO & 1.0 & Puoko-Nui & BG40 & 06:24-11:00 & 20 \\
2011 Jul 4 & MJUO & 1.0 & Puoko-Nui & BG40 & 06:36-14:02 & 20 \\
2011 Jul 6 & MJUO & 1.0 & Puoko-Nui & BG40 & 06:35-09:28 & 20 \\
2011 Jul 27 & MJUO & 1.0 & Puoko-Nui & BG40 & 06:49-12:47 & 20 \\
2011 Aug 1 & MJUO & 1.0 & Puoko-Nui & BG40 & 06:40-12:53 & 20 \\
2011 Aug 2 & MJUO & 1.0 & Puoko-Nui & BG40 & 06:35-10:04 & 20 \\
\enddata
\end{deluxetable}
\normalsize

\begin{deluxetable}{lccc}
\tablewidth{0pt}
\tablecaption{Summary of Best-fit White Dwarf Temperatures}
\tablehead{
\colhead{UT Date} & \colhead{Spectrum} & \colhead{Days past Outburst} & \colhead{Temp ($^{\circ}$K)}}
\startdata
2002 Jan 17 & STIS & pre-outburst & 16,000 \\
2008 Apr 17 & optical & 371 & 25,000 \\
2010 Mar 11 & COS & 1064 & 19,700 \\
2011 Apr 9 & COS & 1458 & 17,300 \\
\enddata
\end{deluxetable}

\clearpage
\begin{deluxetable}{lccc}
\tablewidth{0pt}
\tablecaption{Radial Velocity Fits}
\tablehead{
\colhead{Line} & \colhead{$\gamma$(km s$^{-1}$)} & \colhead{K (km s$^{-1}$)} &
\colhead{$\sigma$ (km s$^{-1}$)} } 
\startdata
CI1463 & 29.3$\pm$0.1 & 7.6$\pm$0.8 & 1.7 \\
SiII1533 & 34.5$\pm$0.1 & 4.8$\pm$1.4 & 2.9 \\
Al1670 & 3.1$\pm$0.2 & 5.4$\pm$1.8 & 3.5 \\
CIV1550 & -95.2$\pm$0.1 & 63.2$\pm$7.4 & 14.7 \\
\enddata
\end{deluxetable}

\clearpage
\begin{deluxetable}{lcccccc}
\tabletypesize{\footnotesize}
\rotate
\tablewidth{0pt}
\tablecaption{Summary of Periodicities}
\tablehead{
\colhead{UT Date} & \colhead{Data} & \colhead{Long P (s)} & \colhead{Amp (mma)} & 
\colhead{Short P (s)} & \colhead{Amp (mma)} & \colhead{3$\sigma$}}
\startdata
2010 Mar 10 & MO & ... & ... & ... & ... & 9.8   \\
2010 Mar 11 & HST & ... & ... & 292.0,289.0,282.1,266.0$\pm$0.1 & 19.5,13.5,18.9,7.7$\pm$0.7 &  4.4 \\
2010 Mar 11 & MO & 2443$\pm$33 & 13.1$\pm$1.9 & 283.6$\pm$0.6  & 9.25$\pm$1.9 &  7.9 \\
2010 Mar 12 & MO & ... & ... & ... & ... & 3.6 \\
2010 Mar 15 & FTS & ... & ... & ... & ... & 8.7 \\
2011 Mar 2 & MJUO & 2477$\pm$44 & 16.0$\pm$2.3 & ... & ... & 8.9 \\
2011 Mar 4 & MJUO & 2525$\pm$55 & 26.8$\pm$2.6 & ... & ... & 10.4 \\
2011 Apr 4 & APO & ... & ... & 276.8$\pm$1.7 & 10.6$\pm$2.1 & 8.8  \\
2011 Apr 8 & APO & 5132,2563$\pm$51,12 & 14.5,16.6$\pm$0.9  & 275.5,298.5$\pm$0.2,0.5 & 9.7,5.7$\pm$0.9 & 3.8  \\
2011 Apr 9 & HST & 2457$\pm$31 & 14.1$\pm$1.4 & 293.3$\pm$0.1 & 47.2$\pm$1.4  & 5.9  \\
2011 May 12 & KPNO & ... & ... & 290,295$\pm$5 & 15,13$\pm$18  & 7.2-12.6\tablenotemark{a} \\
2011 May 13 & KPNO & ... & ... & ... & ... & 9.5 \\
2011 May 24 & KPNO & ... & ... & ... & ... & 10.6 \\
2011 May 25 & KPNO & ... & ... & 329.4,313.6$\pm$2.1,1.4 & 9.7,7.5$\pm$1.8 & 7.1 \\
2011 May 26 & KPNO & ... & ... & 313.75$\pm$0.87 & 16.2$\pm$1.7  & 7.0 \\
2011 Jun 9 & KPNO & ... & ... & 290.8$\pm$2.5 & 8.2$\pm$2.4 & 9.9 \\
2011 Jul 1 & MJUO & ... & ... & 281.0$\pm$0.2 & 13.0$\pm$1.0 & 4.0 \\
2011 Jul 2 & MJUO & ... & ... & 308.5,300.5,292.5$\pm$0.4,0.4,0.2 & 7.8,13.2,6.5$\pm$0.9 & 3.6 \\
2011 Jul 4 & MJUO & 5124$\pm$32 & 15.2$\pm$0.8 & 295.0$\pm$0.2 &  9.5$\pm$0.8 & 3.3 \\ 
2011 Jul 6 & MJUO & 2592$\pm$38 & 11.5$\pm$1.2 & 286.6$\pm$0.3 & 16.5$\pm$1.2 & 5.0 \\
2011 Jul 27 & MJUO & 5149,2542$\pm$74,13 & 11.1,14.5$\pm$1.1 & 275.6$\pm$0.4 & 5.5$\pm$1.0 & 4.5 \\
2011 Aug 1 & MJUO & 2467$\pm$27 & 15.0$\pm$3.2 & ... & ... & 4.0 \\
2011 Aug 2 & MJUO & ... & ... & 323.0$\pm$1.0 & 4.1$\pm$0.9 & 3.8 \\
\enddata
\tablenotetext{a}{Due to the shortness of the light curve, the low frequencies
are unresolved so the shuffling technique gives a larger value than normal. A
lower limit was found by averaging the white noise from 0.004-0.009 Hz.}
\end{deluxetable}
\normalsize

\begin{figure}
\figurenum {1}
\includegraphics[angle=-90,width=7in]{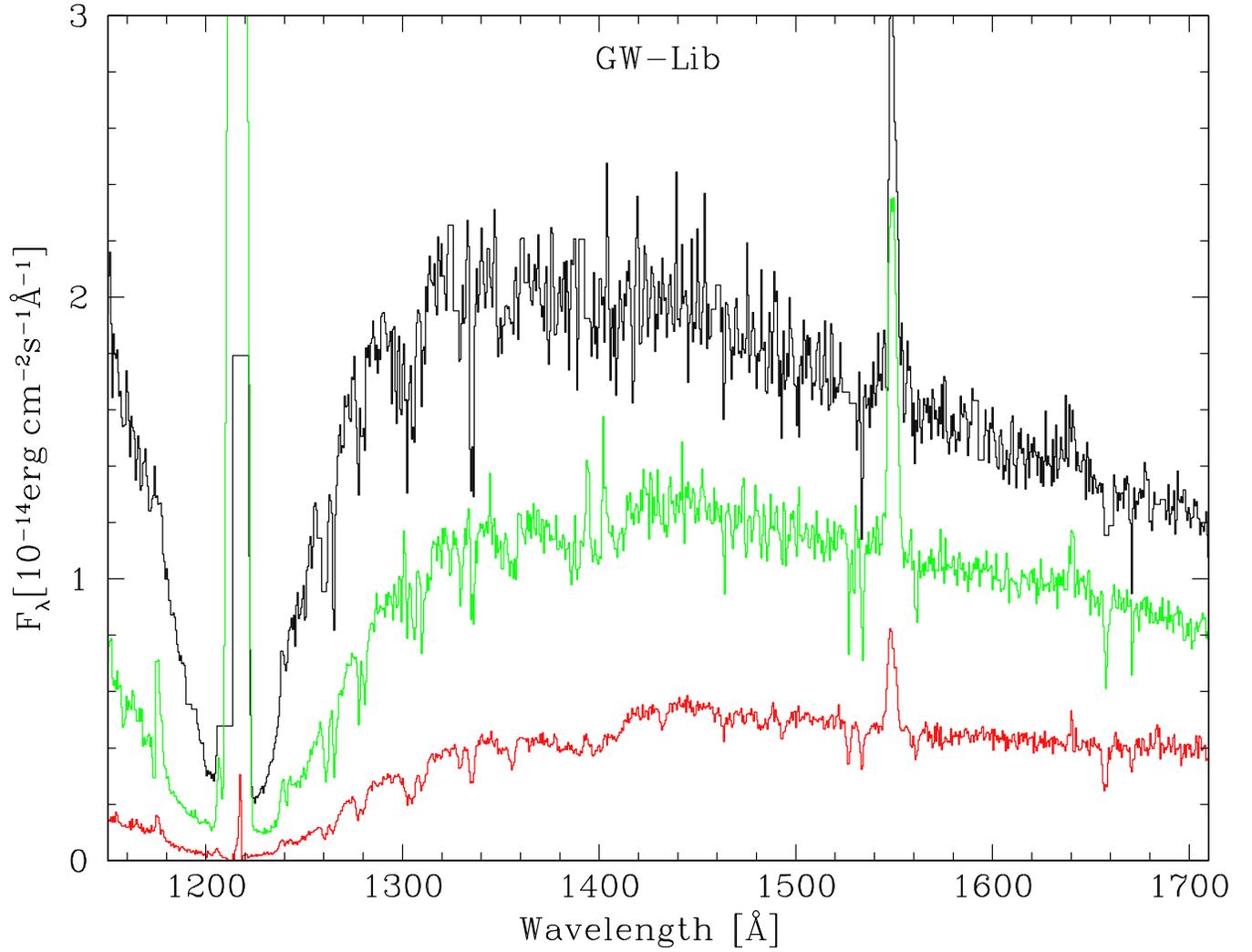}
\caption{Comparison of STIS spectrum at quiescence (bottom) with COS spectra
3 yrs (top) and 4 yrs (middle) past outburst.}
\end{figure}

\clearpage
\begin{figure}
\figurenum{2a}
\includegraphics[angle=-90,width=7in]{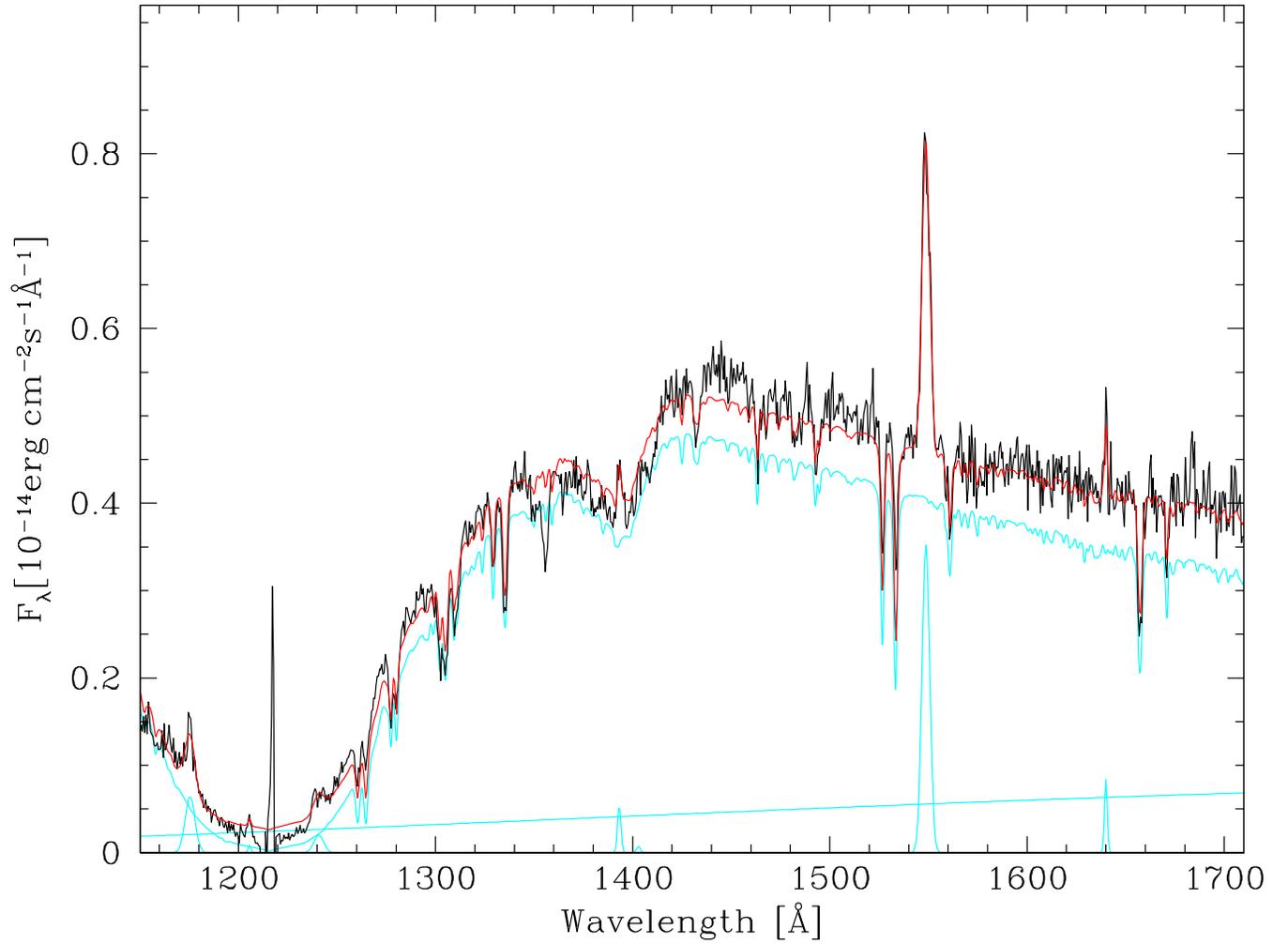}
\caption{STIS spectrum at quiescence fit with a T=16,000K log g=8.7 white dwarf
model, an accretion disk (lower line) and Gaussians at the emission lines.}
\end{figure}

\clearpage
\begin{figure}
\figurenum{2b}
\includegraphics[angle=-90,width=7in]{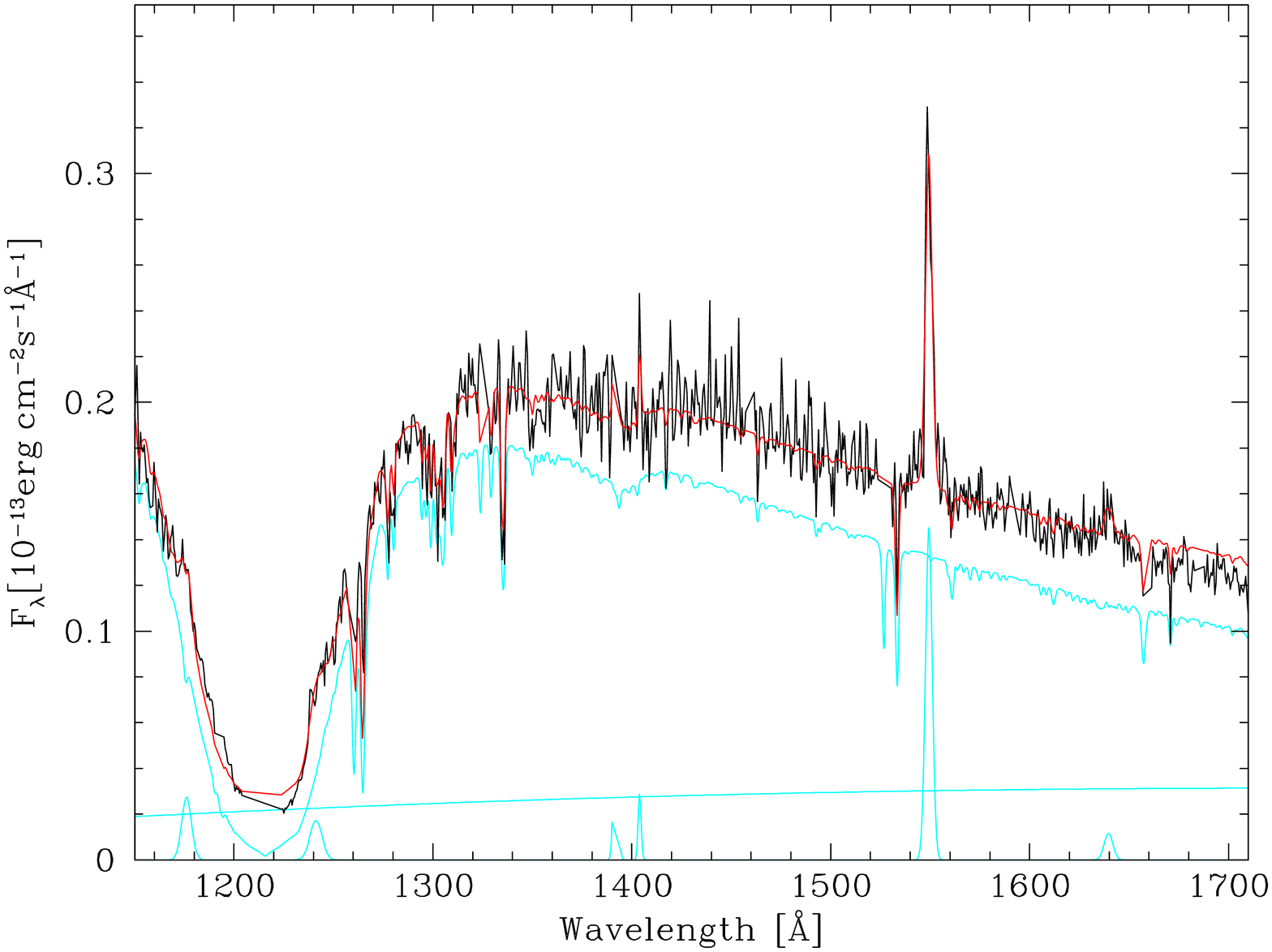}
\caption{2010 COS spectrum 3 yrs past outburst fit with a T=19,700K white dwarf
model.}
\end{figure}

\clearpage
\begin{figure}
\figurenum{2c}
\includegraphics[angle=-90,width=7in]{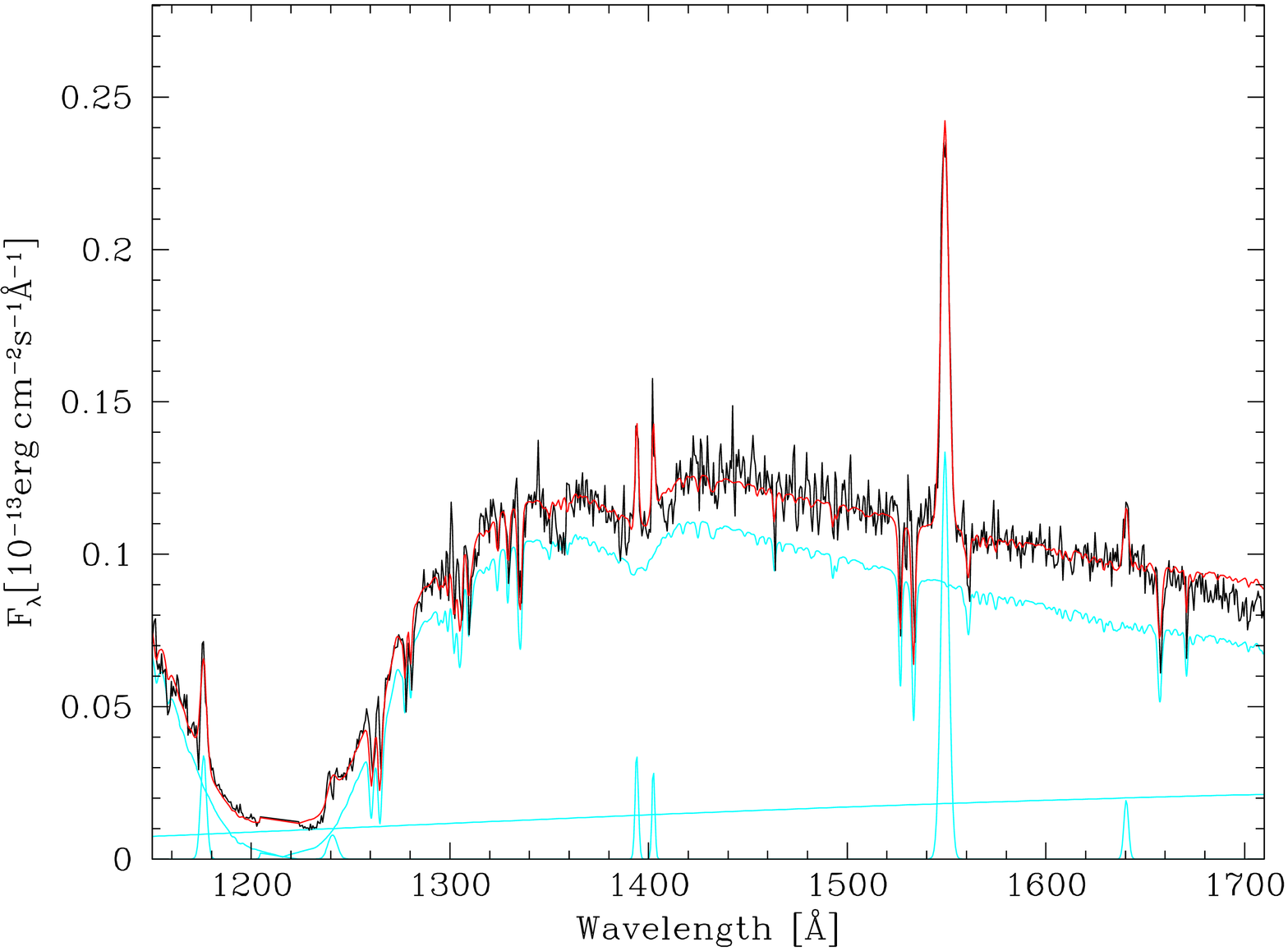}
\caption{2011 COS spectrum 4 yrs past outburst fit with a T=17,300K white dwarf
model.}
\end{figure}

\clearpage
\begin{figure}
\figurenum{3}
\plotone{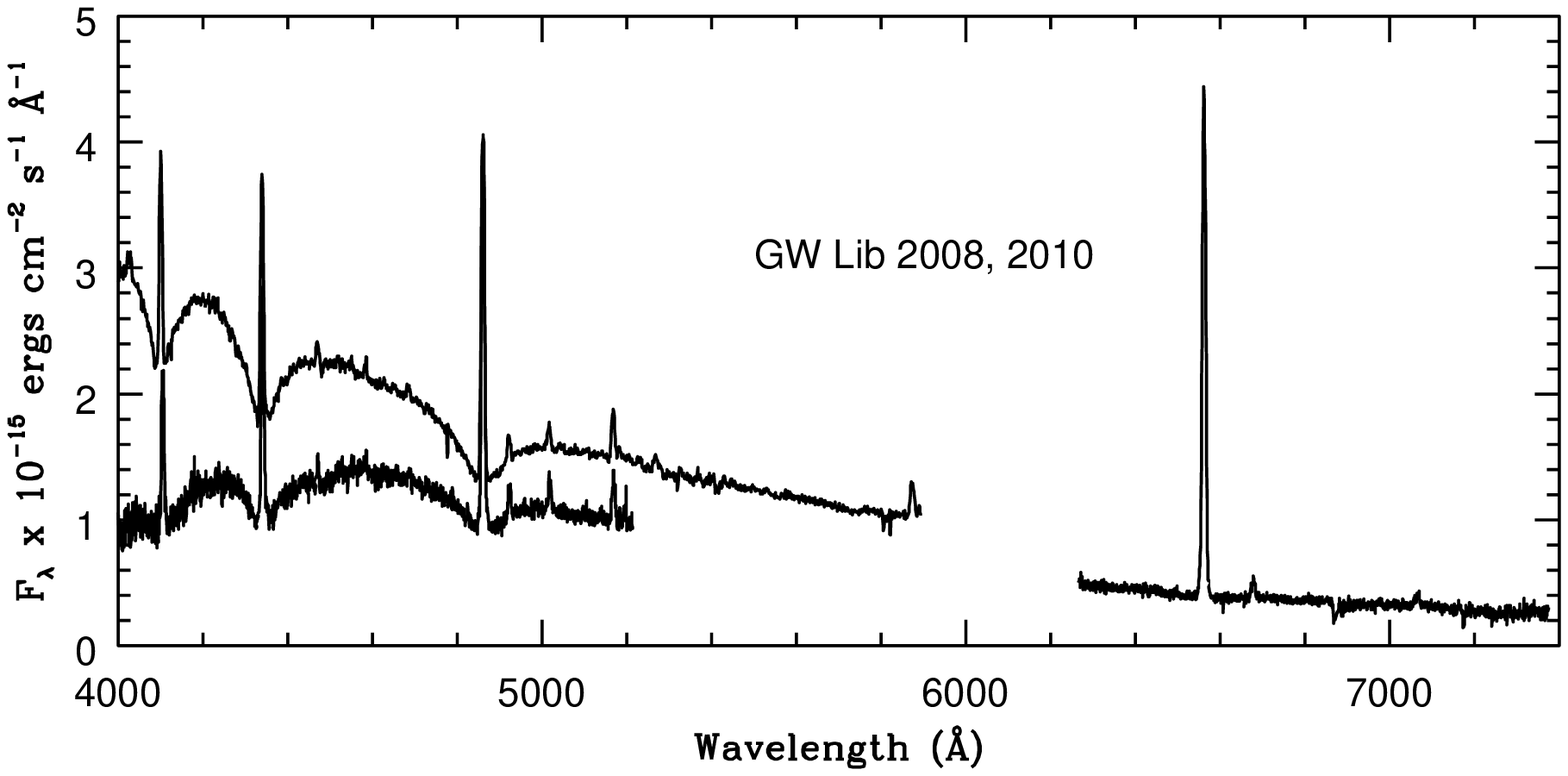}
\caption{Blue optical spectra of GW Lib obtained in 2008 (top left) and 2010 blue
and red spectra (bottom
2 segments).}
\end{figure}

\clearpage
\begin{figure}
\figurenum{4}
\includegraphics[angle=-90,width=7in]{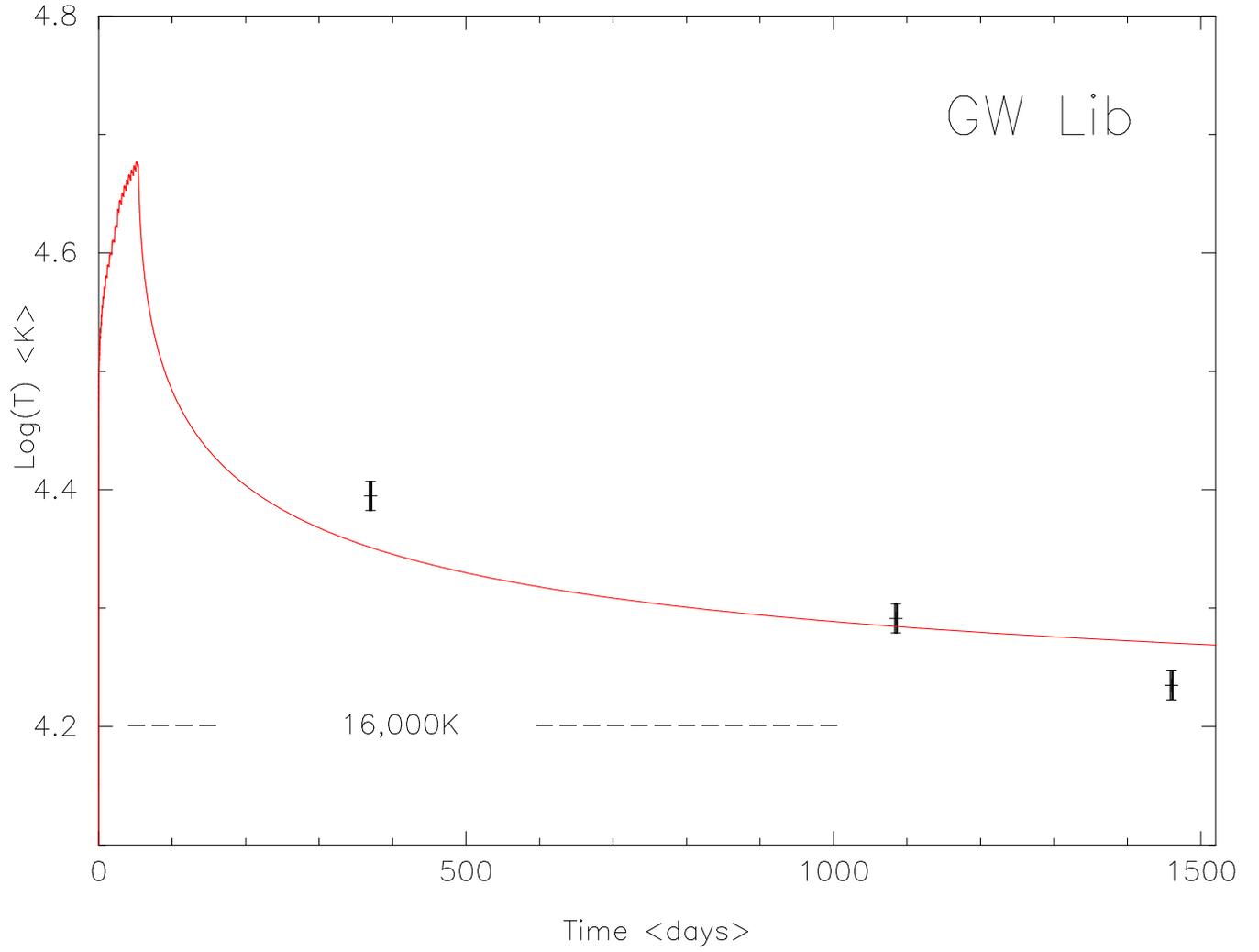}
\caption{Cooling curve computed for an accretion rate of 
5$\times$10$^{-8}$M$_{\odot}$ yr$^{-1}$ compared to the optical (2008) and 
UV (2010, 2011) temperature determinations
at 1-4 yrs post outburst.}
\end{figure}

\clearpage
\begin{figure}
\figurenum{5}
\plotone{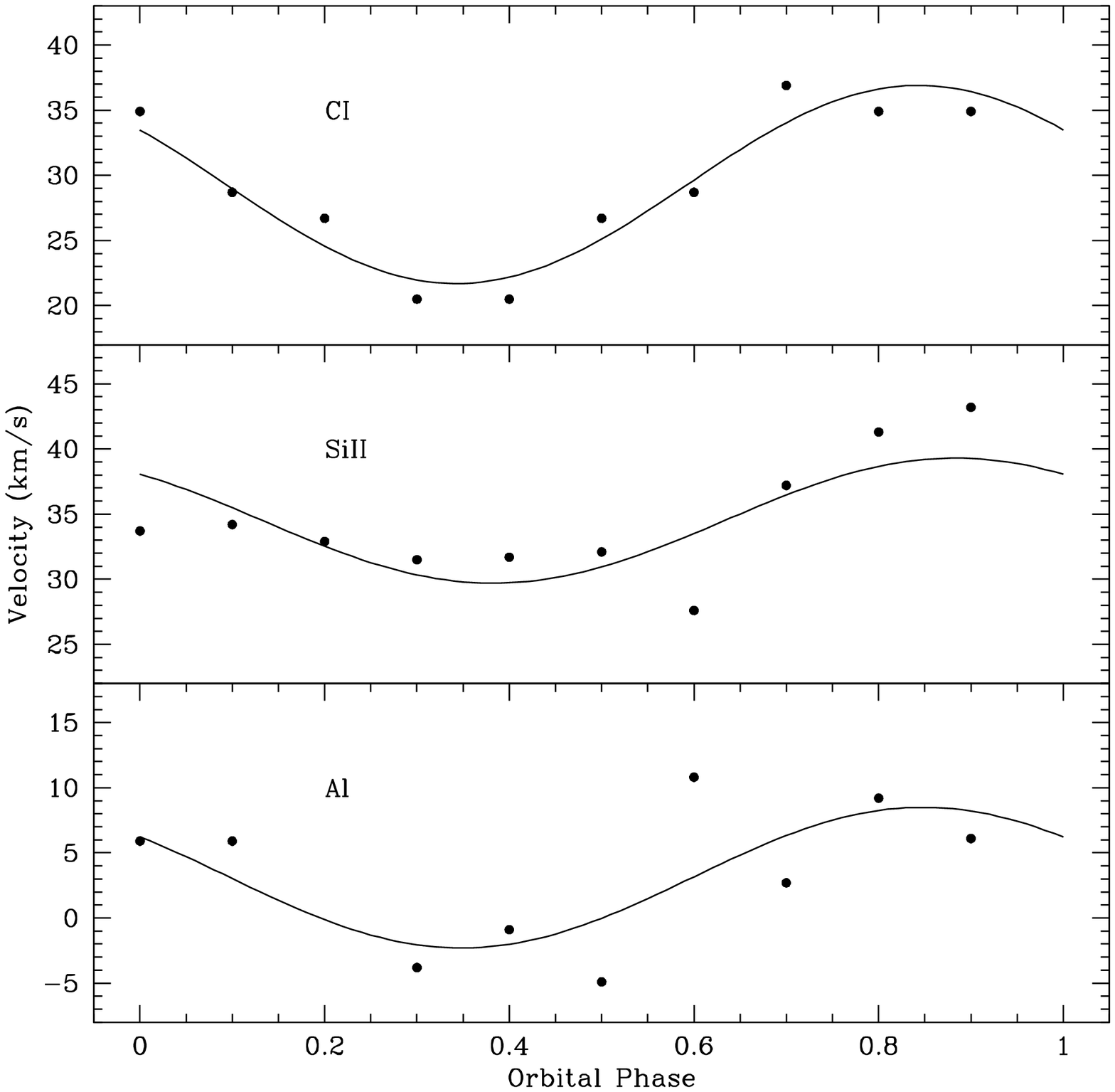}
\caption{Radial velocities and best fit sine-curves for the CI, SiII
and Al absorption lines from the 2010 G160M binned spectra. Statistical error bars
on the fits are 2, 3 and 4 km s$^{-1}$ for CI, SiII and Al (Table 3). Only CI is significant
but others are plotted to show the similar shapes of the fits.}
\end{figure}

\clearpage
\begin{figure}
\figurenum{6}
\plotone{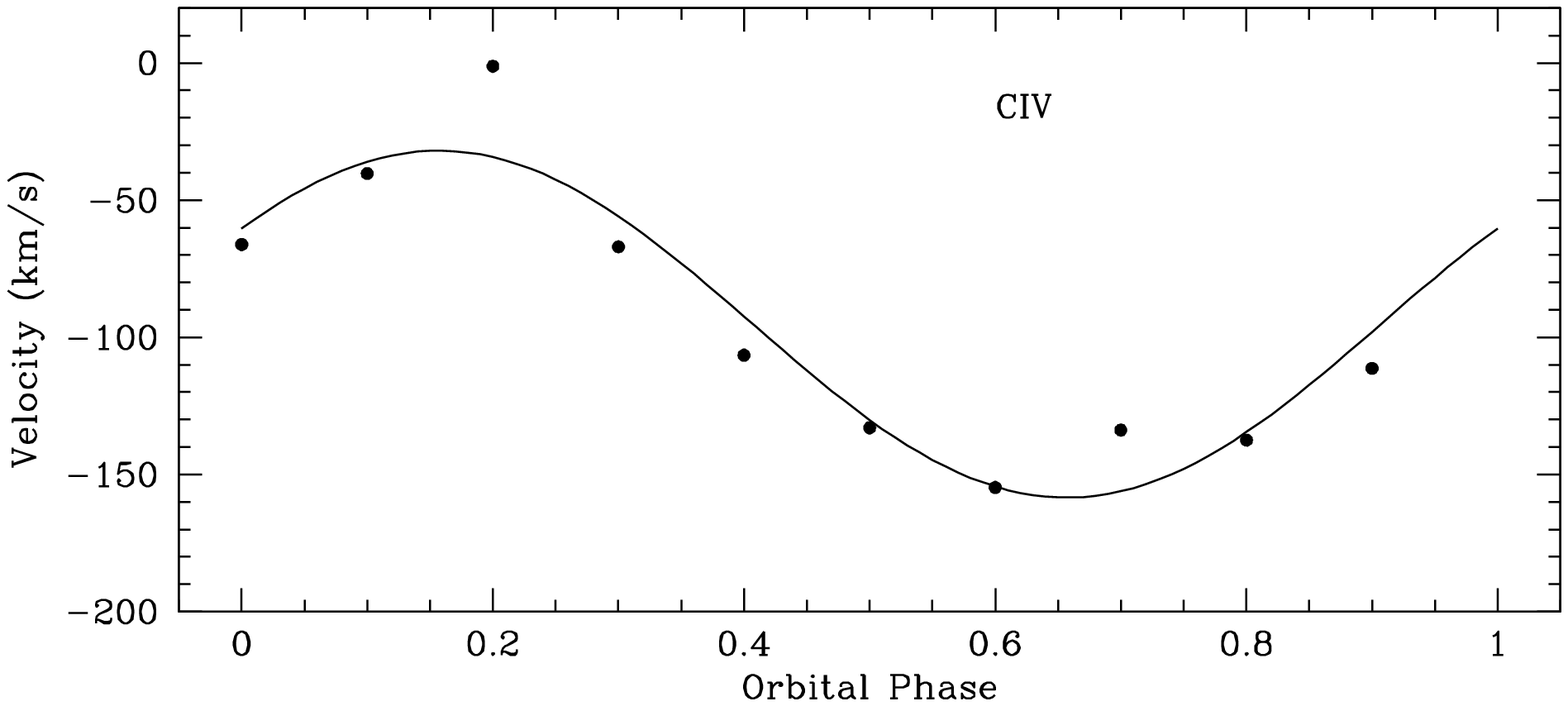}
\caption{Radial velocities and best-fit sine-curve for the CIV emission
line from the 2010 G160M binned spectra. Statistical error of the fit is 15 km s$^{-1}$.}
\end{figure}

\clearpage
\begin{figure}
\figurenum{7}
\includegraphics[angle=-90,width=7in]{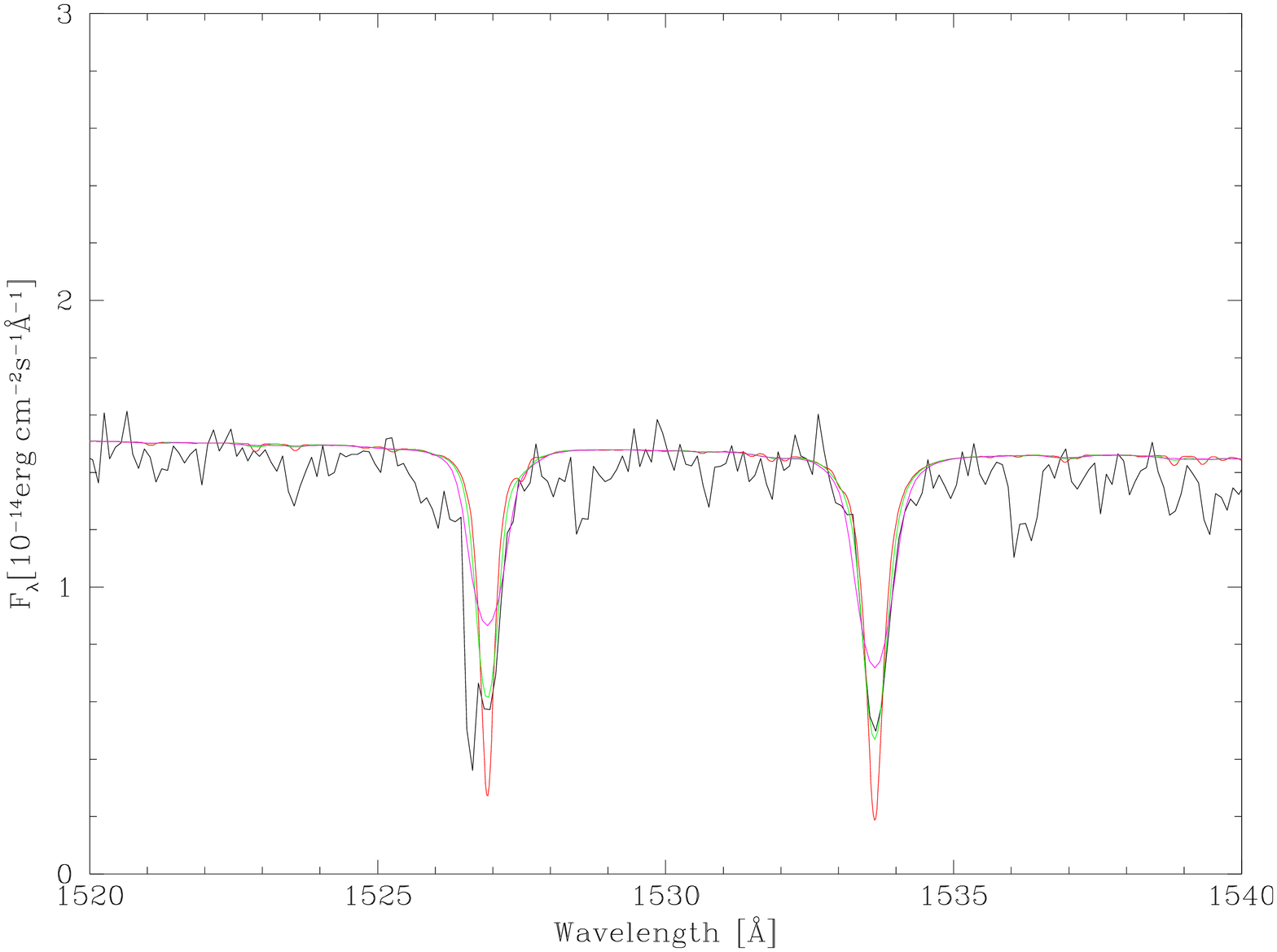}
\caption{G160M spectra fit with 1M$_{\odot}$ white dwarfs with lines
broadened by 20 (red), 50 (green) and 87 (magenta) km s$^{-1}$.}
\end{figure}

\clearpage
\begin{figure}
\figurenum{8}
\plotone{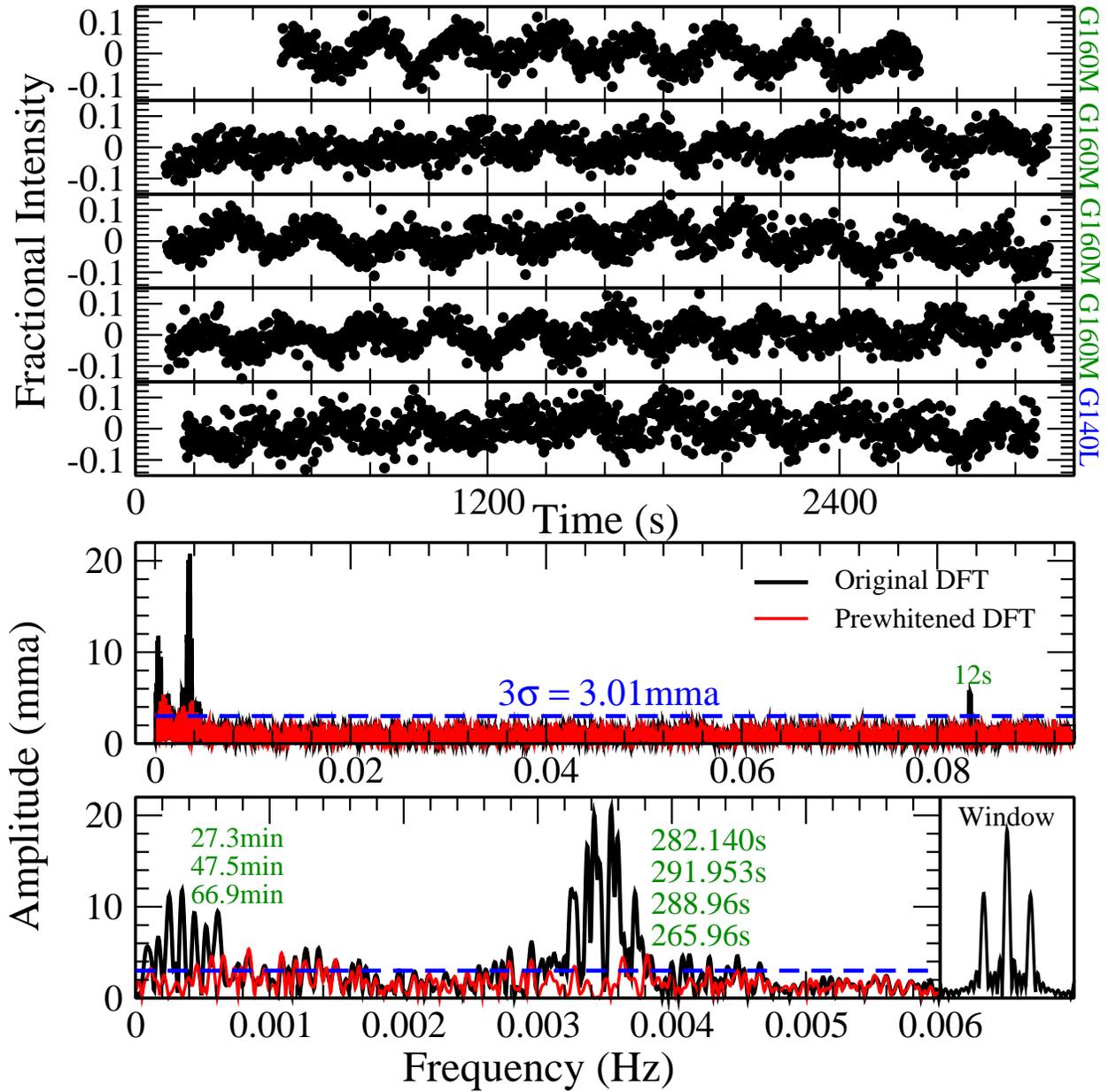}
\caption{Intensity light curves of all 5 orbits of HST 2010 March 11 data
with DFT (middle and expanded at bottom).} 
\end{figure}

\clearpage
\begin{figure}
\figurenum{9}
\plotone{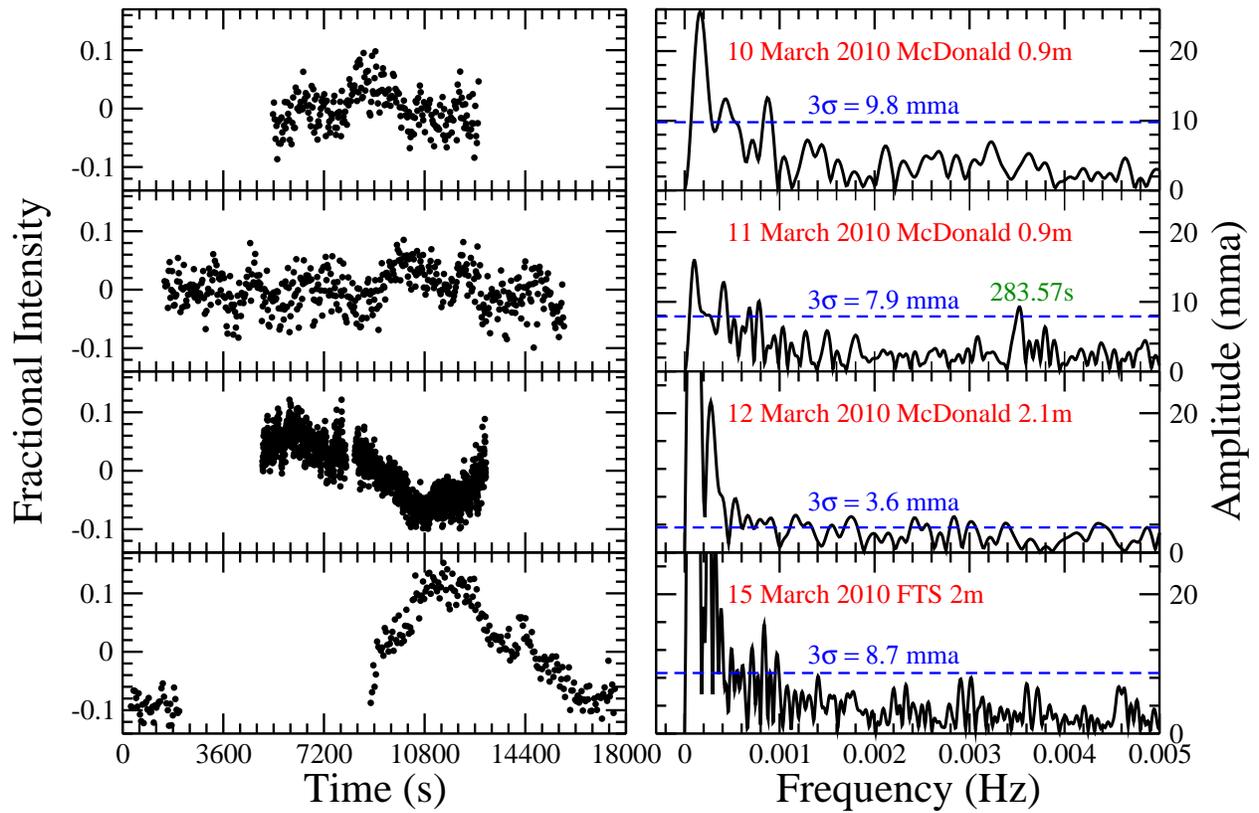}
\caption{Intensity light curves and DFTs for MO and FTS optical data during
2010 March.}
\end{figure}

\clearpage
\begin{figure}
\figurenum{10}
\plotone{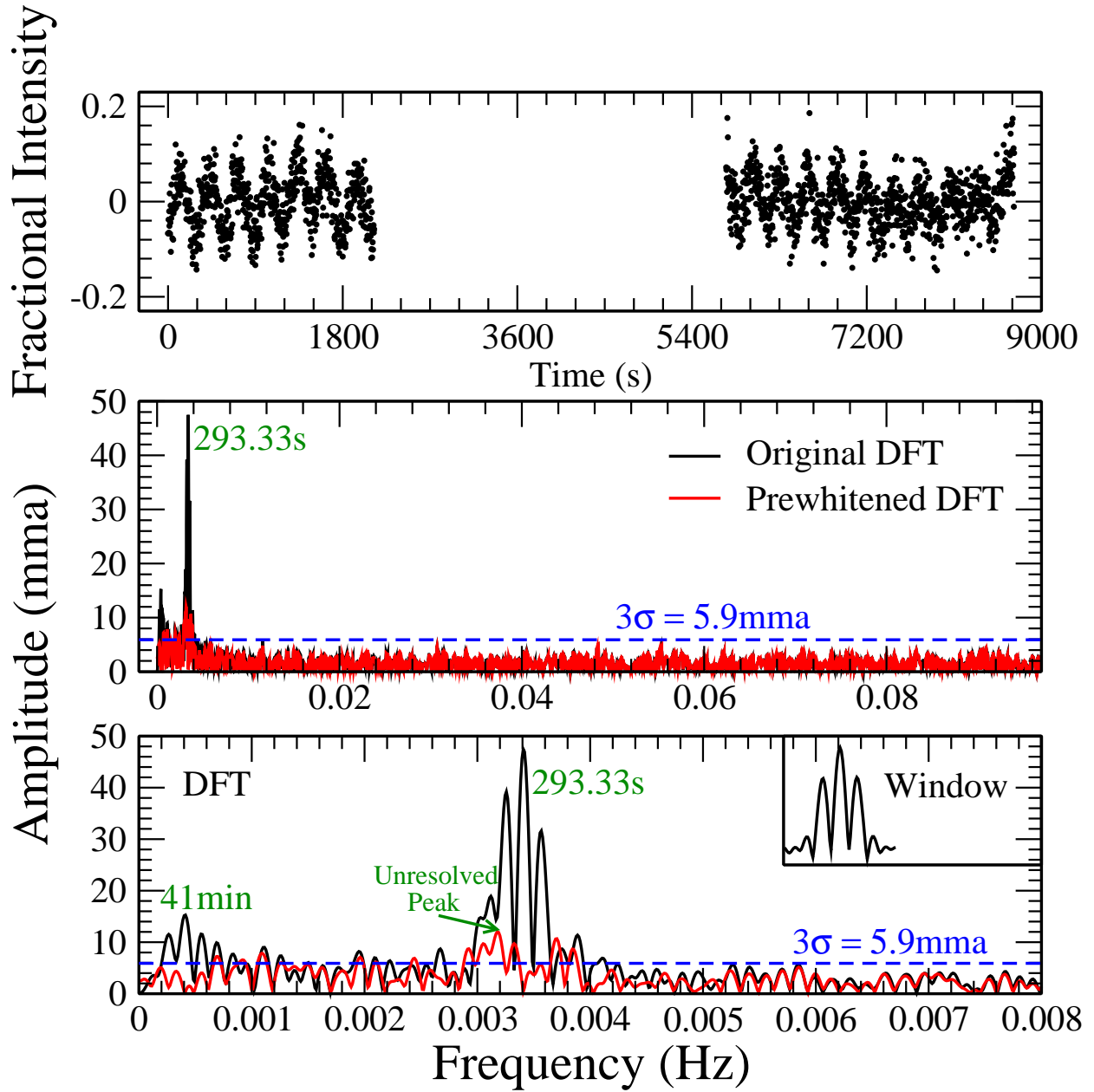}
\caption{Intensity light curve and DFT for 2 orbits of HST data 2011 April 9.}
\end{figure}

\clearpage
\begin{figure}
\figurenum{11}
\plotone{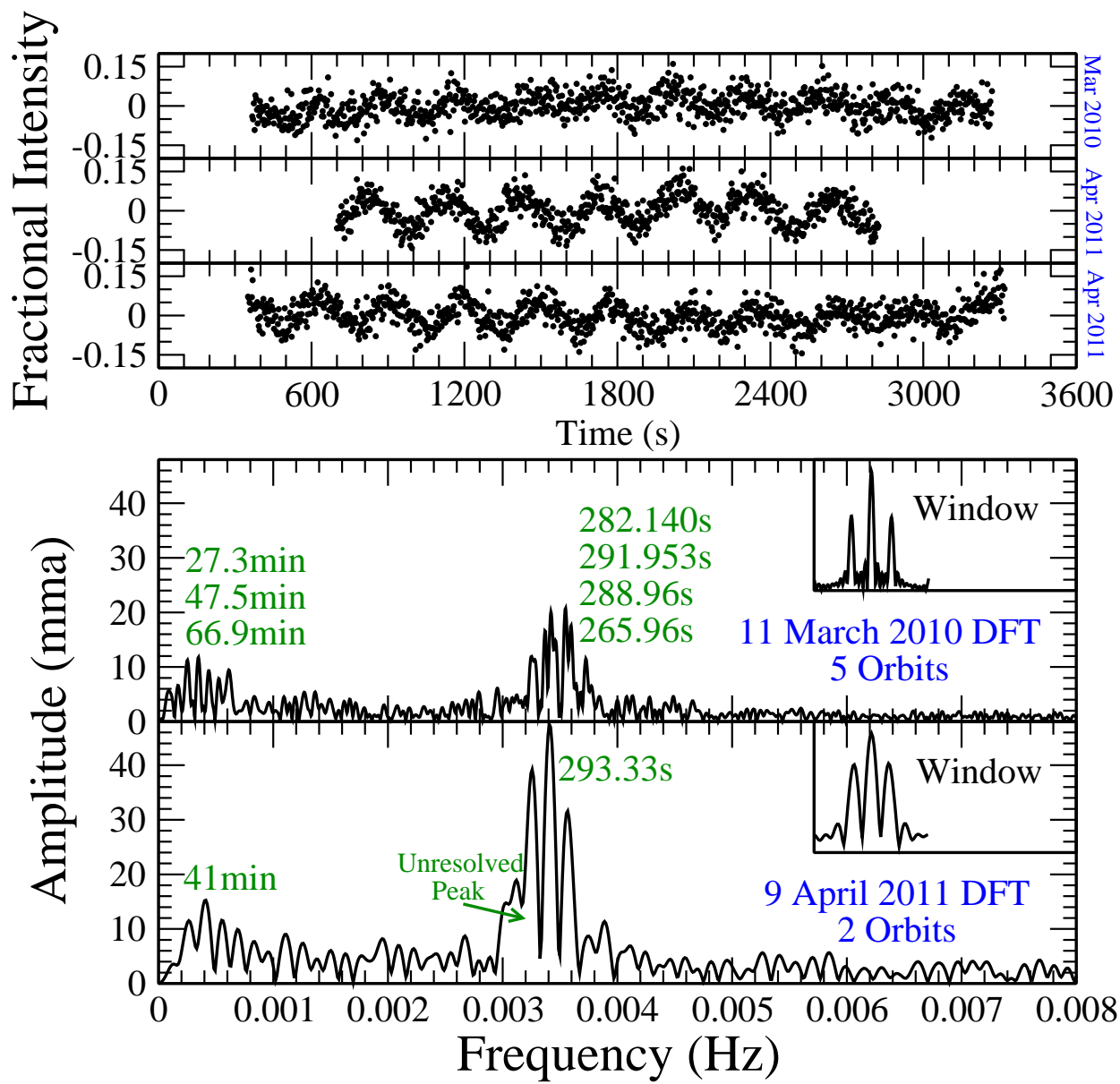}
\caption{Comparison of 2010 and 2011 G140L light curves (top) and DFTs (bottom
where the 2010 DFT is computed from 5 orbits from G160M and G140L and the
2011 data from 2 orbits with G140L).}
\end{figure}

\clearpage
\begin{figure}
\figurenum{12}
\plotone{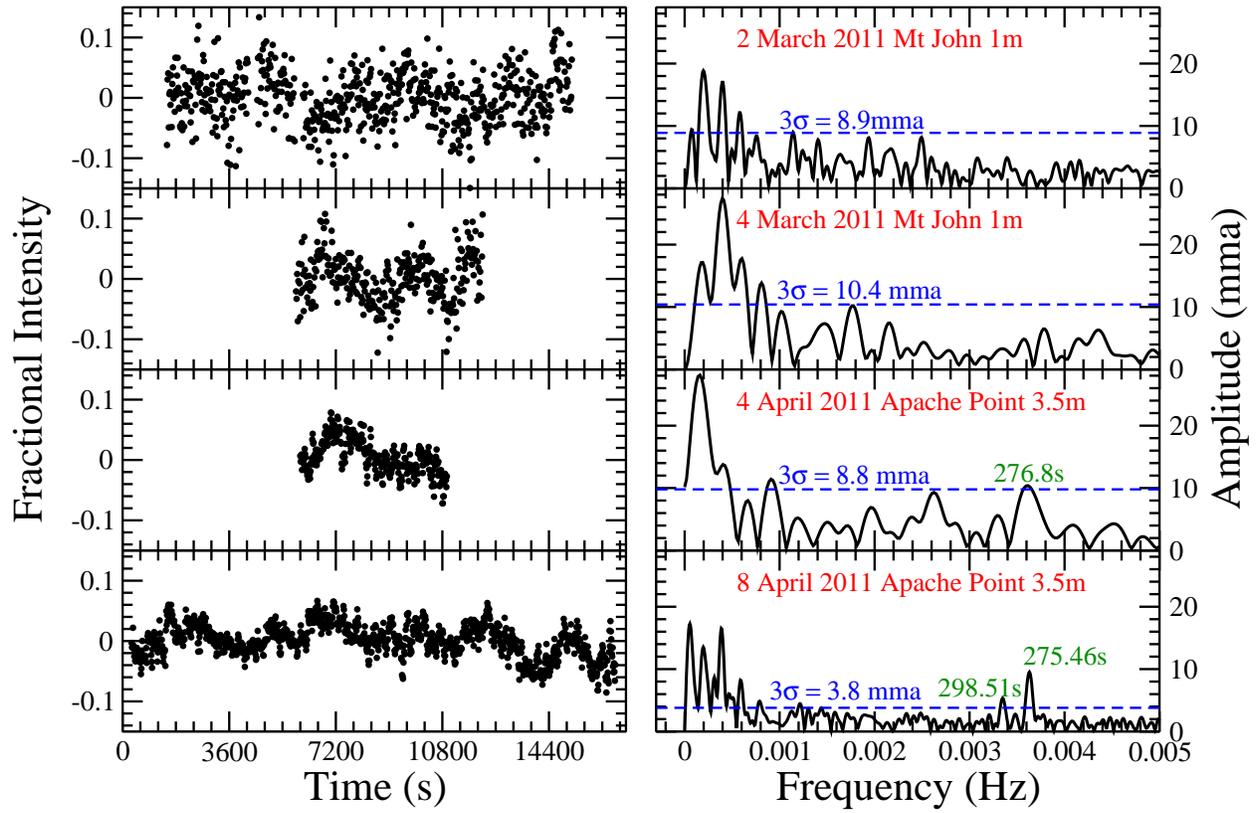}
\caption{Intensity light curves and DFTs of MJUO and APO optical data from 2011
March-April.} 
\end{figure}

\clearpage
\begin{figure}
\figurenum{13}
\plotone{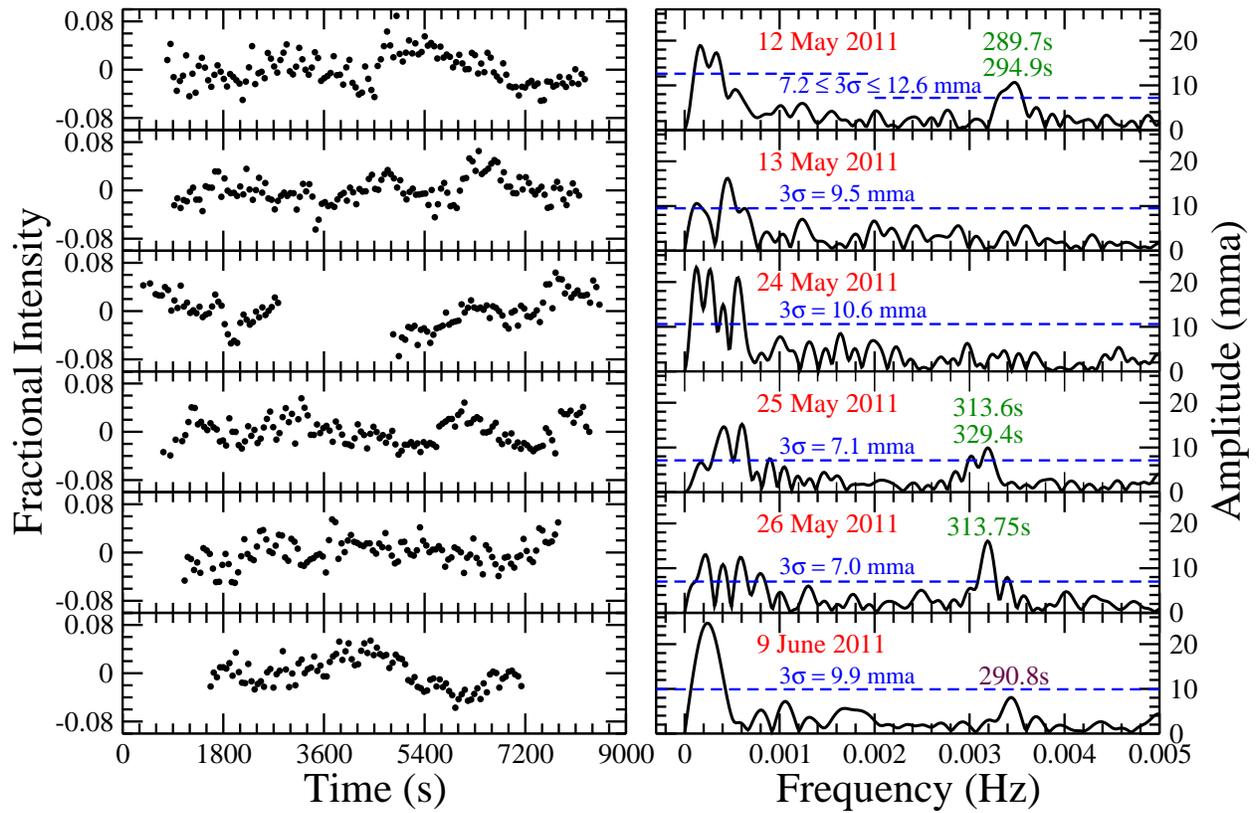}
\caption{Intensity light curves and DFTs from KPNO data during 2011 May-June.}
\end{figure}

\clearpage
\begin{figure}
\figurenum{14}
\plotone{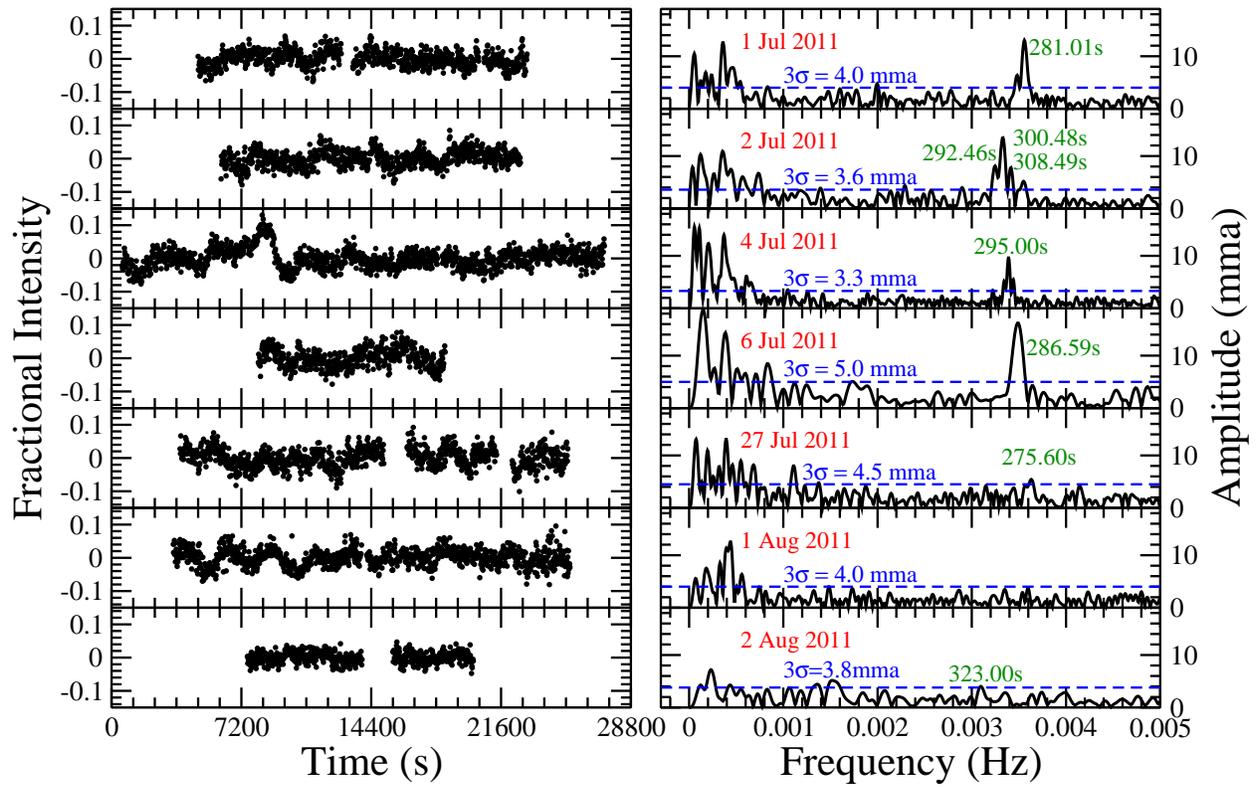}
\caption{Intensity light curves and DFTs from MJUO data during 2011 July-Aug.}
\end{figure}

\clearpage
\begin{figure}
\figurenum{15}
\plotone{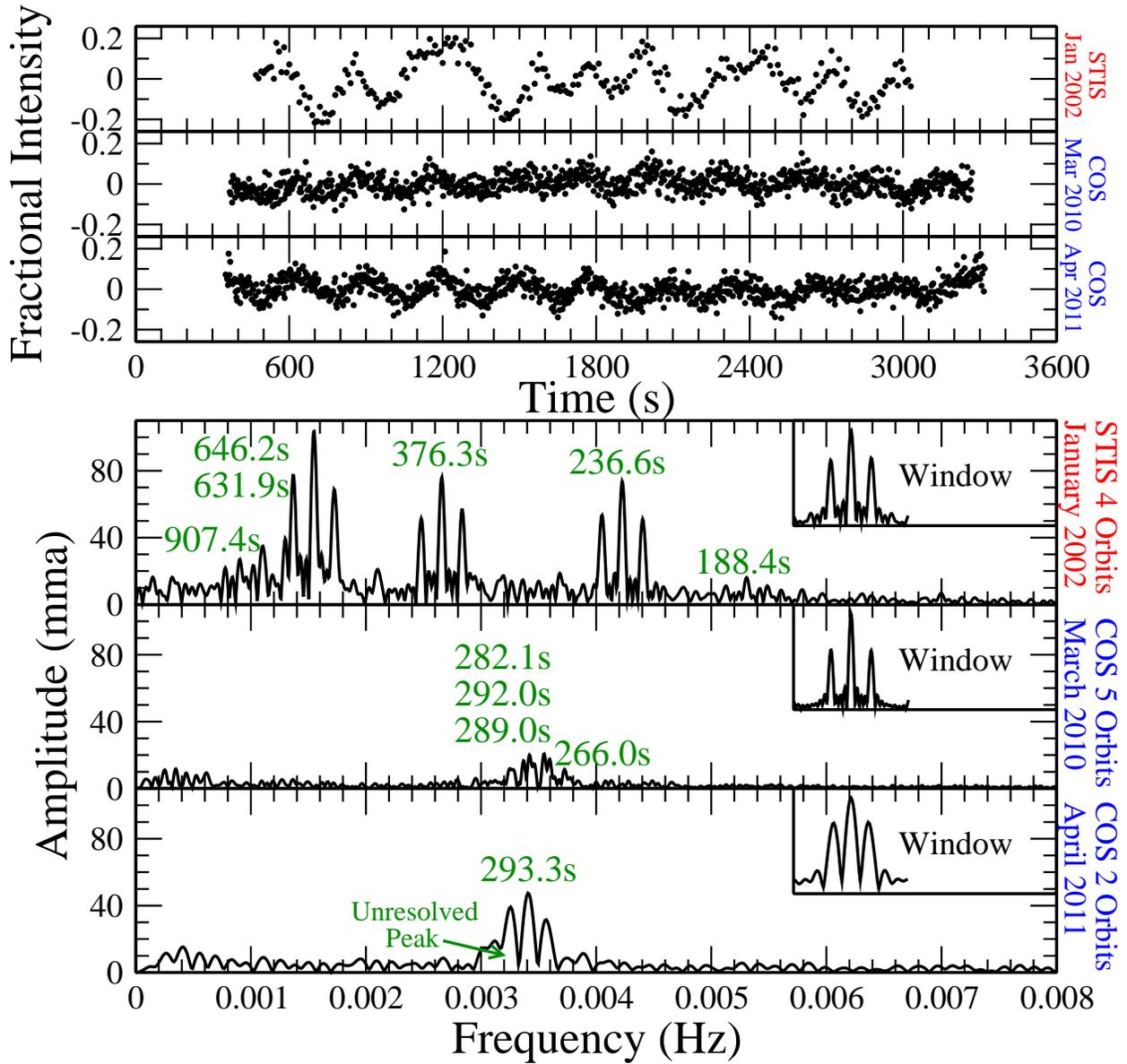}
\caption{Comparison of intensity light curves and DFTs from 2002 STIS quiescent
 data to those from COS 2010 and 2011.}
\end{figure}

\end{document}